\documentclass[review]{elsarticle}
\usepackage{amsmath}
\usepackage{lineno,hyperref}
\modulolinenumbers[200]

\journal{Journal of \LaTeX\ Templates}

\begin{document}

\begin{frontmatter}

\title{Multi-dimensional Jordan chain and Navier-Stokes Equation \tnoteref{mytitlenote}}

\author{B.G.Konopelchenko\fnref{myfootnote}}
\address{Department of Mathematics and Physics, University of Salento, Lecce,Italy}




\begin{abstract}

Multi-dimensional Jordan chain is presented. It is shown that it admits the finite-component reductions to the Navier-Stokes equation and other important equations of continuous media theory
\end{abstract}

\begin{keyword}
\texttt{elsarticle.cls}\sep \LaTeX\sep Elsevier \sep template
\MSC[2010] 00-01\sep  99-00
\end{keyword}

\end{frontmatter}

\linenumbers

\section{Introduction}

\paragraph{} 
The Navier-Stokes equation
\begin{equation}\label{1.1}
\rho\left(\frac{\partial \vec{u}}{\partial t}+\left(\vec{u}\,\,\vec{\nabla}\right)\vec{u}\right)=-\mathrm{grad}\,p+\eta\triangle\vec{u}+\left(\xi+\frac{\eta}{3}\right)\mathrm{grad}(\,\mathrm{div}\,\vec{u}\,)
\end{equation}
where $\vec{u}$, $\rho$, $p$ are velocity, density, pressure and $\eta$, $\xi$ are the viscosity coefficients is one of the basic equations in the theory of continuous media (see e.g. \citep{1,2} ). It arises in the study of a number of phenomena in various branches of physics and has been addressed in large number of researches. It is well-known also that in spite of the apparent simplicity of the Navier-Stokes equation, an analysis of its properties is a tough problem.

In the present paper the relation between the Navier-Stokes equation and the multi-dimensional Jordan chain is estabilished. The n-dimensional Jordan chain is an infinite set of quasi-linear equations of the form
\begin{equation}\label{1.2}
\frac{\partial u_{ln+i}}{\partial t}+\sum_{k=1}^nu_k\frac{\partial u_{ln+i}}{\partial x_k}+\frac{\partial u_{(l+1)n+i}}{\partial x_i}=0 \qquad, \qquad i=1,...,n;l=0,1,2,... \,.
\end{equation}
It is shown that under the constraint
\begin{equation}\label{1.3}
-\rho\frac{\partial u_{n+i}}{\partial x_i}=-\frac{\partial \rho}{\partial x_k}+\eta\sum_{k=1}^n\frac{\partial^2u_i}{\partial x_k^2}+\left(\xi+\frac{\eta}{3}\right)\frac{\partial}{\partial x_i}\left(\,\sum_{k=1}^n\frac{\partial u_k}{\partial x_k}\right)\,\, ,\,\, i=1,...,n
\end{equation}
the Jordan chain (\ref{1.2}) is reduced to the n-dimensional Navier-Stokes equation (equation (\ref{1.1}) at $n=3$). Other important equations like multi-dimensional Euler equation and multi-dimensional Burgers equation also are the appropriate reductions of the Jordan chain (\ref{1.2}).

It is demonstrated that a class of solutions of the Jordan chain (\ref{1.2}) and its reductions is provided implicitly by the hodograph type equations
\begin{align}\label{1.4}
& x_i=u_it+f_i(u_1,u_2,...)\qquad ,\qquad i=1,...,n\\\nonumber
&0=t+f_k(u_1,u_2,...)\qquad ,\qquad k=n+1,n+2,...
\end{align}
with the certain constraints on the functions $f_i(u_1,u_2,...)$, $i=1,2...$.

The paper is organized as follows. In section 2 the N-component n-dimensional Jordan system is constructed. The Jordan chain (\ref{1.2}) is its formal limit at $N\rightarrow\infty$. Reductions of the Jordan chain to the n-dimensional Navier-Stokes equation and other equations are considered in section 3.



\section{Multi-dimensional Jordan system}
First we will construct the "N-component" n-dimensional Jordan system with arbitrary N and n. The hodograph system ($N\geq 2$)
\begin{equation}\label{2.1}
x_i=u_it+f_i(u_1,...,u_{Nn})\qquad ,\qquad i=1,...,n,
\end{equation}
\begin{equation}\label{2.2}
0=t+f_i(u_1,...,u_{Nn})\qquad ,\qquad i=n+1,...,Nn
\end{equation}
where $f_i(u_1,...,u_{Nn})$, $i=1,..,Nn$ are functions of $Nn$ variables is our starting point.

Differentiating these equations with respect to $x_k$, $k=1,...,n$, one gets
\begin{equation}\label{2.3}
\delta_{ik}=\sum_{l=1}^{Nn}\left(t\delta_{il}+\frac{\partial f_i}{\partial u_l}\right)\frac{\partial u_l}{\partial x_k}\qquad ,\qquad i,k=1,...,n\,
\end{equation}
and
\begin{equation}\label{2.4}
0=\sum_{l=1}^{Nn}\frac{\partial f_i}{\partial u_l}\frac{\partial u_l}{\partial x_k}\qquad ,\qquad i=n+1,...,Nn;k=1,...,n
\end{equation}
where $\delta_{ik}$ is the Kroneker symbol. Introducing $Nn\times Nn$ matrices $\omega$ and $A$ defines by $\omega_{ik}=\delta_{ik}$, $i,k=1,...,n$, $\omega_{ik}=0$, $i=n+1,...,Nn$ and/or $k=n+1,...,Nn$ and
\begin{equation}\label{2.5}
A_{ik}=t\omega_{ik}+\frac{\partial f_i}{\partial u_k}\qquad ,\qquad i,k=1,...,Nn,
\end{equation}
one rewrites the relations (\ref{2.3}),(\ref{2.4}) as 
\begin{equation}\label{2.6}
\omega_{ik}=\sum_{l=1}^{Nn}A_{il}\frac{\partial u_l}{\partial x_k}\qquad ,\qquad i=1,...,Nn;k=1,...,n.
\end{equation}
Then, differentiating (\ref{2.1}),(\ref{2.2}) with respect to $t$, one obtains
\begin{equation}\label{2.7}
-V_i=\sum_{l=1}^{Nn}A_{il}\frac{\partial u_l}{\partial t}\qquad ,\qquad i=1,...,Nn,
\end{equation}
where $V_i=u_i$, $i=1,...,n$ and $V_i=1$, $i=n+1,...,Nn$. Hence, if $\det A\neq 0$, one has
\begin{equation}\label{2.8}
\frac{\partial u_i}{\partial x_k}=\sum_{l=1}^{Nn}\left(A^{-1}\right)_{il}\omega_{lk}\qquad ,\qquad i=1,...,Nn;k=1,...,n
\end{equation}
and
\begin{equation}\label{2.9}
\frac{\partial u_i}{\partial t}=-\sum_{l=1}^{Nn}\left(A^{-1}\right)_{il}V_l\qquad ,\qquad i=1,...,Nn.
\end{equation}
Due to specific form of $\omega$ and $V$ these relations are equivalent to the following
\begin{equation}\label{2.10}
\frac{\partial u_i}{\partial x_k}=\left(A^{-1}\right)_{ik}\qquad ,\qquad i=1,...,Nn;k=1,...,n
\end{equation}
and
\begin{equation}\label{2.11}
\frac{\partial u_i}{\partial t}=-\sum_{k=1}^{n}\left(A^{-1}\right)_{ik}u_k-\sum_{m=n+1}^{Nn}\left(A^{-1}\right)_{im}\qquad ,\qquad i=1,...,Nn.
\end{equation}
The relations (\ref{2.10}),(\ref{2.11}) imply that a solution $u_i(i=1,...,Nn)$ of the hodograph equations (\ref{2.1}),(\ref{2.2}) for any given functions $f_1,...,f_{Nn}$ obey the system of equations
\begin{equation}\label{2.12}
\frac{\partial u_i}{\partial t}+\sum_{k=1}^{n}u_k\frac{\partial u_i}{\partial x_k}+\sum_{m=n+1}^{Nn}\left(A^{-1}\right)_{im}=0\qquad ,\qquad i=1,...,Nn.
\end{equation}

Now one imposes the constraints
\begin{equation}\label{2.13}
\sum_{m=n+1}^{Nn}\left(A^{-1}\right)_{ln+i,m}=\frac{\partial u_{(l+1)n+i}}{\partial x_i}\qquad , \qquad i=1,...,n;l=0,1,...,N-2,
\end{equation}
\begin{equation}\label{2.14}
\sum_{m=n+1}^{Nn}\left(A^{-1}\right)_{(N-1)n+i,m}=0\qquad ,\qquad i=1,...,n.
\end{equation}
Under such constraints the system (\ref{2.12}) assumes the form
\begin{equation}\label{2.15}
\frac{\partial u_{ln+i}}{\partial t}+\sum_{k=1}^nu_k\frac{\partial u_{ln+i}}{\partial x_k}+\frac{\partial u_{(l+1)n+i}}{\partial x_i}=0\qquad , \qquad i=1,...,n;l=0,1,...,N-2,
\end{equation}
\begin{equation}\label{2.16}
\frac{\partial u_{(N-1)n+i}}{\partial t}+\sum_{k=1}^nu_k\frac{\partial u_{(N-1)n+i}}{\partial x_k}=0\qquad ,\qquad i=1,...,n.
\end{equation}
It is the "N-component" n-dimensional Jordan system.

The system (\ref{2.15}),(\ref{2.16}) is the multi-dimensional generalization of the one dimensional ($n=1$) Jordan system introduced in \citep{3}. The system similiar to (\ref{2.15}),(\ref{2.16}) at $N=2$ has been constructed in \citep{4} in a slightly different way. The case $N=1$ corresponds to the homogeneous Euler equation
\begin{equation}\label{2.17}
\frac{\partial u_i}{\partial t}+\sum_{k=1}^{n}u_k\frac{\partial u_i}{\partial x_k}=0\qquad ,\qquad i=1,...,n.
\end{equation}
The applicability of the hodograph equations (\ref{2.1}) and their Lagrangian version to the system (\ref{2.17}) has been demonstrated in \citep{5,6}and \citep{7,8}.

Due to the relations (\ref{2.10}) the constraints (\ref{2.13}),(\ref{2.14}) are equivalent to the following constraints on the elements of the matrix $A^{-1}$
\begin{equation}\label{2.18}
\sum_{m=n+1}^{Nn}\left(A^{-1}\right)_{ln+i,m}=\left(A^{-1}\right)_{(l+1)n+i,i}\qquad ,\qquad i=1,...,n;l=0,1,...,N-2,
\end{equation}
\begin{equation}
\sum_{m=n+1}^{Nn}\left(A^{-1}\right)_{(N-1)n+i,m}=0\qquad ,\qquad i=1,...,n.\nonumber
\end{equation}
These constraints represent themselves the system of $Nn$ nonlinear partial differential equations for $Nn$ functions $f_i(u_1,...,u_{Nn})$. Any solution of this system provides us implicitly, via the hodograph equations (\ref{2.1}),(\ref{2.2}), with the solutions of the Jordan system (\ref{2.15}),(\ref{2.16}).

Solutions of the Jordan system can be also constructed using the hodograph type equations
\begin{align}\label{2.19}
&x_i=u_it+f_i\left(u_1,...,u_{Nn}\right)\qquad ,\qquad i=1,...,n,\nonumber\\
& 0=t+f_{n+1}\left(u_1,...,u_{Nn}\right)\\
&0=g_k\left(u_1,...,u_{Nn}\right)\qquad ,\qquad k=n+2,...,Nn\nonumber
\end{align}
which obviously are equivalent to the system of equations (\ref{2.1}),(\ref{2.2}). At $n=1$ the hodograph equations (\ref{2.19}) has been used in \citep{4}.

Similiar to the one-dimensional case \citep{9} an imbedding of equations (\ref{2.17}) into the Jordan system (\ref{2.15}),(\ref{2.16}) is a way to regularize the blow-ups of derivatives and gradient catastrophes for the homogeneous Euler equation. Indeed, due to the relations (\ref{2.10}),(\ref{2.11}) the first level blow-ups for the equations (\ref{2.17}) occurs on the hypersurface defines by the equation 
\begin{equation}
\det A(N=1)=0\label{2.20}
\end{equation}
where
\begin{equation}
A_{ik}(N=1)=t\delta_{ik}+\frac{\partial x_i}{\partial u_k}\qquad ,\qquad i,k=1,...,n.\nonumber
\end{equation}
After the imbedding into the "2-component" Jordan system (\ref{2.15}),(\ref{2.16}) ($N=2$) the condition (\ref{2.20}) is no more critical and the corresponding first level blow-ups disappear. First level blow-ups for the "2-component" Jordan system which occur on the hypersurface $\det A(N=2)=0$ are regularized in the imbedding into the "3-component" Jordan system and so on. In detail such type of regularization of the blow-ups for the n-dimensional Euler equation (\ref{2.17}) will be considered in a separate paper.
\section{Multi-dimensional Jordan chain and reduction to Navier-Stokes equation}
Now let us consider the formal limit $N\rightarrow\infty$ of the system (\ref{2.15}),(\ref{2.16}). In this limit it becomes an infinite set of equations of the form
\begin{equation}\label{3.1}
\frac{\partial u_{ln+i}}{\partial t}+\sum_{k=1}^{n}u_k\frac{\partial u_{ln+i}}{\partial x_k}+\frac{\partial u_{(l+1)n+i}}{\partial x_i}=0\qquad ,\qquad i=1,...,n;l=0,1,2,... \,.
\end{equation}
We will refer to this set of equations as the n-dimensional Jordan chain. In the one-dimensional case ($n=1$) it has been introduced in different ways in the papers \citep{3,10,11}.

Similiar to the one-dimensional case the chain (\ref{3.1}) can be extended and used in various directions. Here we will consider only some of its finite-component reductions.

The first reduction is generated by the constraint
\begin{equation}\label{3.2}
\frac{\partial u_{n+i}}{\partial x_i}=-\frac{1}{\rho}\left(-\frac{\partial p}{\partial x_i}+\eta\Delta u_i+\left(\xi+\frac{\eta}{3}\right)\frac{\partial}{\partial x_i}\left(\sum_{k=1}^n\frac{\partial u_k}{\partial x_k}\right)\right)\qquad ,\qquad i=1,...,n
\end{equation}
where the density $\rho$ obeys the continuity equation
\begin{equation}\label{3.3}
\frac{\partial \rho}{\partial t}+\sum_{i=1}^n\frac{\partial}{\partial x_i}\left(\rho u_i\right)=0.
\end{equation}
Here $p$ is the pressure, $\Delta\equiv\sum_{k=1}^n\frac{\partial^2}{\partial x_k^2}$ and $\eta$, $\xi$ are arbitrary constants. Under this constraint the first equation ($l=0$) of the Jordan chain (\ref{3.1}) assumes the form
\begin{equation}\label{3.4}
\rho\left(\frac{\partial u_i}{\partial t}+\sum_{k=1}^nu_k\frac{\partial u_i}{\partial x_k}\right)=-\frac{\partial p}{\partial x_i}+\eta\Delta u_i+\left(\xi+\frac{\eta}{3}\right)\frac{\partial}{\partial x_i}\left(\sum_{k=1}^n\frac{\partial u_k}{\partial x_k}\right)\, ,\, i=1,...,n.
\end{equation}
The subsequent equations with $l=1,2,3,...$, i.e.
\begin{equation}\label{3.5}
\frac{\partial u_{(l+1)n+i}}{\partial x_i}=-\frac{\partial u_{ln+i}}{\partial t}-\sum_{k=1}^nu_k\frac{\partial u_{ln+i}}{\partial x_k}\qquad ,\qquad i=1,2,...,n;l=1,2,3,...
\end{equation}
represent themselves the recurrent relation for the calculation of all derivatives $\frac{\partial u_{(l+1)n+i}}{\partial x_i}$, $i=1,...,n$, $l=1,2,3,...$ in terms of $\frac{\partial u_{n+i}}{\partial x_i}$, $i=1,...,n$ given by (\ref{3.2}). The inclusion of equation (\ref{3.3}) apparently does not produce additional constraints.

The system (\ref{3.4}),(\ref{3.3}) is the n-dimensional version of the classical basic equation (\ref{1.1}),(\ref{3.3}) ($n=3$) of the theory of continuous media \citep{1,2}. 

Constraint (\ref{3.2}) is equivalent to the following constraint for the functions $f_i$:
\begin{align}\label{3.6}
&\rho\left(A^{-1}\right)_{n+i,i}+\eta\sum_{k=1}^n\sum_{m=1}^\infty \frac{\partial \left(A^{-1}\right)_{ik}}{\partial u_m}\left(A^{-1}\right)_{mk}+\\\nonumber
&+\left(\xi+\frac{\eta}{3}\right)\sum_{k=1}^n\sum_{m=1}^\infty \frac{\partial \left(A^{-1}\right)_{kk}}{\partial u_m}\left(A^{-1}\right)_{mi}=\frac{\partial p}{\partial x_i}\qquad ,\qquad i=1,...,n.
\end{align}
Here $\rho$ obeys the equation (\ref{3.3}) and $p(x)$ is an arbitrary function.

This constraint together with the constraints (\ref{2.18}) at $N\rightarrow\infty$ characterizes those functions $f_i$, $i=1,2,...$ for which the hodograph equations (\ref{2.1}),(\ref{2.2}) at $N\rightarrow\infty$ provide us with the solutions of the n-dimensional Navier-Stokes equation (\ref{3.4}).

Constraint (\ref{3.2}) with $\eta=\xi=0$ gives rise to the classical Euler equation which together with the continuity equation (\ref{3.3}) describes the motion of inviscid compressible fluid \citep{1,2}.

In the approximation of incompressible inviscid fluid ($\rho=const$) for which \citep{1}
\begin{equation}\label{3.7}
\Delta p=-\rho\sum_{i,k=1}^{n}\frac{\partial u_k}{\partial x_i}\frac{\partial u_i}{\partial x_k}
\end{equation}
the constraint (\ref{3.2}) assumes the form
\begin{equation}\label{3.8}
\sum_{k=1}^n\frac{\partial^2u_{n+k}}{\partial x_k^2}=-\sum_{i,k=1}^n\frac{\partial u_k}{\partial x_i}\frac{\partial u_i}{\partial x_k}.
\end{equation}
The corresponding constraint on $A_{ik}$ is given by
\begin{equation}\label{3.9}
\sum_{k=1}^n\sum_{m=1}^\infty \frac{\partial \left(A^{-1}\right)_{m+k,k}}{\partial u_m}\left(A^{-1}\right)_{mk}=-\, \sum_{i,k=1}^n\left(A^{-1}\right)_{ki}\left(A^{-1}\right)_{ik}.
\end{equation}

Further, imposing the constraint
\begin{equation}\label{3.10}
\frac{\partial u_{n+i}}{\partial x_i}=\nu\Delta u_i\qquad ,\qquad i=1,...,n
\end{equation}
or equivalently
\begin{equation}
\left(A^{-1}\right)_{n+i,i}=\nu\sum_{k=1}^n\sum_{m=1}^\infty \frac{\partial \left(A^{-1}\right)_{ik}}{\partial u_m}\left(A^{-1}\right)_{mk}\qquad ,\qquad i=1,...,n,\nonumber
\end{equation}
one reduces the Jordan chain to the n-dimensional Burgers equation
\begin{equation}
\frac{\partial u_i}{\partial t}+\sum_{k=1}^nu_k\frac{\partial u_i}{\partial x_k}+\nu\Delta u_i=0\qquad ,\qquad i=1,...,n.\label{3.11}
\end{equation}
This equation also has appeared in the study of various phenomena (see e.g. \citep{12}).

At $n=1$ such a reduction has been considered in \citep{10,4}.

Another type of finite-component reductions of the Jordan chain (\ref{2.1}) is associated with the parametrization
\begin{equation}\label{3.12}
u_{ln+i}=\frac{1}{ln+1}\sum_{\alpha=1}^M\epsilon_{i\alpha}\left(V_{i\alpha}\right)^{ln+1}\qquad ,\qquad i=1,...,n;l=0,1,2,...,
\end{equation}
where $\epsilon_{i\alpha}$ are arbitrary constants and $M$ is arbitrary integer ($\geq 1$). Under such ansatz equations (\ref{3.1}) become 
\begin{align}\label{3.13}
&\sum_{\alpha=1}^M\epsilon_{i\alpha}\left(V_{i\alpha}\right)^{ln}\left\lbrace\frac{\partial V_{i\alpha}}{\partial t}+\sum_{k=1}^n\sum_{\beta=1}^M\epsilon_{k\beta}V_{k\beta}\frac{\partial V_{i\alpha}}{\partial x_k}+\left(V_{i\alpha}\right)^n\frac{\partial V_{i\alpha}}{\partial x_i}\right\rbrace=0\, ,\, \\&i=1,...,n;l=0,1,2,...\,\,\,.\nonumber
\end{align}
Hence, the variables $V_{i\alpha}$ obey the system of equations
\begin{equation}\label{3.14}
\frac{\partial V_{i\alpha}}{\partial t}+\sum_{k=1}^n\lambda_{i\alpha ,k}\frac{\partial V_{i\alpha}}{\partial x_k}=0\qquad ,\qquad i=1,...,n;\alpha=1,...,M
\end{equation}
where
\begin{equation}
\lambda_{i\alpha ,k}=\sum_{\beta=1}^M\epsilon_{k\beta}V_{k\beta}+\delta_{ik}\left(V_{i\alpha}\right)^n\qquad ,\qquad k,i=1,...,n;\alpha=1,...,M.\nonumber
\end{equation}
This n-dimensional hydrodynamic type system is the multi-dimensional generalization of the one-dimensional $\epsilon$-sistem discussed in \citep{11}.

It is noted that the system (\ref{3.14}) apparently is not a reduction of the matrix n-dimensional homogeneous Euler equation considered in \citep{13}.

The system (\ref{3.14}) can be viewed as the source of solutions for the Jordan chain (\ref{3.1}) and reduced equations presented above. Indeed, let us take a solution $\lbrace V_{i\alpha}, i=1,...,n;\alpha=1,...,M \rbrace$ of the system (\ref{3.14}) and introduce the power sum variables $u_{ln+i}$, $i=1,...,n$;$l=0,1,...$ defined by (\ref{3.12}). Then the relation (\ref{3.13}) implies that these $u_{ln+i}$ are the solutions of the Jordan chain (\ref{3.1}). Varying $M$ and $\epsilon_{i\alpha}$, one obtains different classes of solutions for the Jordan chain.

Now let us consider the system (\ref{3.14}) accompanied by the differential constraints 
\begin{align}\label{3.15}
& \rho\sum_{\alpha=1}^M\epsilon_{i\alpha}\left( V_{i\alpha}\right)^n\frac{\partial V_{i\alpha}}{\partial x_i}+\eta\sum_{\alpha=1}^M\epsilon_{i\alpha}\Delta V_{i\alpha}+\\
&+\left(\xi+\frac{\eta}{3}\right)\sum_{k=1}^n\sum_{\alpha=1}^M\epsilon_{k\alpha}\frac{\partial^2 V_{k\alpha}}{\partial x_i\partial x_k}=\frac{\partial p}{\partial x_i}\,\, ,\,\, i=1,...,n,\nonumber
\end{align} 
\begin{equation}\label{3.16}
\frac{\partial \rho}{\partial t}+\sum_{i=1}^n\sum_{\alpha=1}^M\frac{\partial}{\partial x_i}\left(\rho\epsilon_{i\alpha}V_{i\alpha}\right)=0.
\end{equation}
Then, due to the relation (\ref{3.2}) solutions of the system (\ref{3.14})-(\ref{3.16}) are the solutions of the Navier-Stokes equation (\ref{3.4}) with $u_i=\sum_{\alpha=1}^M\epsilon_{i\alpha}V_{i\alpha}$, $i=1,...,n$.

Classes of solutions of the multi-dimensional Burgers equation (\ref{3.11}) are obtainable by the formula $u_i=\sum_{\alpha=1}^M\epsilon_{i\alpha}V_{i\alpha}$, $i=1,...,n$ where $V_{i\alpha}$ are the solutions of the system composed by equation (\ref{3.14}) and constraint
\begin{equation}\label{3.17}
\sum_{\alpha=1}^M\epsilon_{i\alpha}\left( V_{i\alpha}\right)^n \frac{\partial V_{i\alpha}}{\partial x_i}=\nu \sum_{\alpha=1}^M\epsilon_{i\alpha}\Delta V_{i\alpha}\qquad ,\qquad i=1,...,n.
\end{equation}

The constraints (\ref{3.15}),(\ref{3.17}) are rather cumbersome. So the efficiency of such a procedure evidently depends on the degree of solvability of the systems of equations given above. This problem and that of applicability of the hodograph equations method to the system (\ref{3.14}) definitely worth investigation.


\section*{References}
\bibliographystyle{elsarticle-num}
\bibliography{mybibfile}

\end{document}